# GUIDE: Graphical User Interface Data for Execution

*Rajat Chawla*    *Adarsh Jha*    *Muskaan Kumar*

*Mukunda NS*    *Ishaan Bhola*

SuperAGI Research

## Abstract

In this paper, we introduce GUIDE, a novel dataset tailored for the advancement of Multimodal Large Language Model (MLLM) applications, particularly focusing on Robotic Process Automation (RPA) use cases. Our dataset encompasses diverse data from various websites including Apollo(62.67%), Gmail(3.43%), Calendar(10.98%) and Canva(22.92%). Each data entry includes an image, a task description, the last action taken, CoT and the next action to be performed along with grounding information of where the action needs to be executed. The data is collected using our in-house advanced annotation tool NEXTAG (Next Action Grounding and Annotation Tool). The data is adapted for multiple OS, browsers and display types. It is collected by multiple annotators to capture the variation of design and the way person uses a website.

Through this dataset, we aim to facilitate research and development in the realm of LLMs for graphical user interfaces, particularly in tasks related to RPA. The dataset's multi-platform nature and coverage of diverse websites enable the exploration of cross-interface capabilities in automation tasks. We believe that our dataset will serve as a valuable resource for advancing the capabilities of multi-platform LLMs in practical applications, fostering innovation in the field of automation and natural language understanding. Using GUIDE, we build V-Zen, the first RPA model to automate multiple websites using our in-House Automation tool AUTONODE[1].

**Explore Guide**: `https://huggingface.co/datasets/SuperAGI/GUIDE`

## 1. Introduction

Robotic Process Automation (RPA) has emerged as an innovative tool that enables the automation of repetitive and rule-based tasks in various domains such as finance, healthcare, and customer service. As organizations strive to improve efficiency and scalability, RPA solutions have become integral components of digital transformation strategies. However, the current state of RPA is largely dependent on pre-defined scripts and rules, limiting its adaptability and responsiveness to dynamic interfaces and workflows. The burgeoning field of artificial intelligence (AI) presents an opportunity to enhance RPA with cognitive capabilities, enabling systems to interpret and interact with Graphical User Interfaces (GUIs) intelligently, much like a human operator would. In the realm of Robotic Process Automation (RPA), the automation of graphical user interfaces (GUIs) stands as a significant challenge that combines the intricacies of understanding the interface layout with the complexities of task execution sequences and an end to end workflow imitating user interaction with the platform.

Traditional datasets in this domain often concentrate on isolated aspects of the task, such as object recognition or sequence prediction, without integrating the crucial dimension of action grounding – the spatial localization of actionable elements on the GUI. To bridge this gap, we introduce the GUIDE (Graphical User Interface Data for Execution) dataset, designed to revolutionize the training of RPA models through data-driven learning. The GUIDE dataset is distinct from existing resources in its comprehensive amalgamation of image data, task descriptions, action histories, chains of thought (COT), and spatial grounding of actions, collected from a variety of web applications and services. This dataset provides a foundation for training multi-platform Large Language Models (LLMs) that can seamlessly predict and execute tasks within a GUI context, adding a layer of semantic understanding that supersedes the capabilities of traditional RPA tools. In this paper, we delve into the creation and structure of GUIDE, illustrating its potential to enable LLMs to perform next-action prediction and spatial grounding of actions across various web platforms. Moreover, GUIDE is thoughtfully curated to represent a realistic scope of websites, covering diverse services like Apollo, Gmail, Calendar, Canva, Twitter, and Instagram. Each website contributes to the dataset's heterogeneity in design and interaction patterns, offering a robust testing ground for RPA models.

With the development of GUIDE, we aim to foster innovation in RPA research, improving both the efficiency and intelligence of automated systems. This introduction will encompass the motivation for creating GUIDE, its relevance in the current RPA landscape, the methodology used for data collection and annotation, and the potential applications of this dataset in advancing the field of RPA. We also introduce V-Zen, a pioneering RPA model that leverages GUIDE to demonstrate exceptional performance in multi-interface, cross-platform automation tasks using our In-House Automation tool, AUTONODE [1]. This paper elucidates the characteristics of GUIDE, outlines our methodologies in gathering and annotating the data, demonstrates the application of V-Zen, and discusses the implications this holds for the future of GUI-based task automation in an increasingly digital world. In the following sections, we will discuss related work that inspired GUIDE, elaborate on the dataset collection and annotation process, present our experiments with V-Zen, and explore the implications of our findings for future research and development in intelligent automation.

The rest of the paper is organized as follows. Section 2 presents the literature review. Section 3 presents the stages of data collection. Section 4 presents the task difficulty. Section 5 shows the data augmentation. Section 6 presents the limitations. Section 7 presents the experimental overview. Section 8 presents the conclusion.

## 2. Literature Survey

The development of Large Language Models (LLMs) has been significantly influenced by the evolution of multimodal datasets and models. This progress in artificial intelligence emphasizes the integration of various data types, including text, images, videos, and audio, to enhance the capabilities of machines in comprehension and generation. In the early stages, pioneering datasets such as Visual Genome[2], MS COCO[3], TextCaps[4], VQA[5], OK-VQA[6], AOK-VQA[7], GQA[8], OCR-VQA[9], TextVQA[10], and VIzWiz[11] played a crucial role in establishing a fundamental connection between visual content and textual annotations. This connection set the foundation for models to address tasks like image captioning and visual question answering (VQA)[9]. These initial efforts paved the way for more sophisticated datasets that aimed to tackle the complexities of human language and vision interactions. For example, the CLEVR[12] dataset focused on compositional language and elementary visual reasoning, providing a structured environment to evaluate models' reasoning abilities. Concurrently, the VQA[9] datasets emerged as a standard for assessing a model's proficiency in understanding and answering questions based on visual input. These foundational datasets played a vital role in fostering the development of early multimodal models, which in turn laid the groundwork for subsequent innovations in the field.

As the field progressed, increased attention to the specificity of tasks and domains resulted in the creation of more targeted datasets and models, specifically addressing challenges in RPA[13] and GUI interactions. Notably, datasets like UiPath's[14] Public Dataset provided real-world automation scenarios, facilitating the development of models tailored for RPA[13] tasks. The introduction of models like LXMERT[15], which specialize in cross-modal attention mechanisms, played a crucial role in achieving a nuanced understanding necessary for navigating the complexities of graphical user interfaces. Additionally, advancements in models such as TAPAS[16], which focus on table-based question answering, and PixelBERT[17], which align image pixels with text segments, demonstrate significant progress in efficient and context-aware multimodal interpretation.

During the evaluation of grounding capabilities in AI models, including GPT-4[18], Gemini[19], and Claude[20], an interesting observation emerged regarding the disparity between grounding proficiency and predictive accuracy. Despite showcasing competence in predicting future actions or generating coherent responses, none of these models managed to strike a satisfactory balance between effective grounding and accurate prediction of the coordinates. This indicates that while these systems may excel in one aspect, either grounding or prediction, they fall short of achieving proficiency in both simultaneously. This discrepancy highlights the ongoing challenge of integrating robust grounding mechanisms into AI architectures, emphasizing the need for further refinement to attain comprehensive understanding and responsive capability within intelligent systems. GUIDE aims to tackle this challenge and endeavors to excel in both domains.

The GUIDE dataset, which includes detailed annotations covering images, task descriptions, actions, and grounding information from various web platforms such as Apollo, Gmail, and Canva, extends the current knowledge base. Its goal is to support research in automating interactions across different operating systems, browsers, and display formats, bridging the divide between LLM capabilities and real-world applications like AU. By integrating knowledge from established models and their interaction with diverse datasets, GUIDE establishes itself as a crucial tool for enhancing MLLM applications, opening up new possibilities for innovation in automation and natural language comprehension.

The GUIDE dataset has the potential to drive significant advancements in MLLMs[21], particularly in the field of Robotic Process Automation (RPA) and interactions with graphical user interfaces (GUIs). It expands the scope of multimodal datasets and models, drawing inspiration from recent models like Llava[22], BLIVA[23], Veagle[24] and RoBERTa[25] These models have demonstrated impressive abilities in understanding and generating content across multiple modalities. The introduction of GUIDE serves the purpose of not only bridging the current gaps but also expanding the possibilities of what multimodal LLMs can achieve practically. By meticulously annotating data from various web platforms with a focus on actions, tasks, and contextual grounding, GUIDE represents a comprehensive resource that reflects the diverse nature of human-digital interface interactions. These detailed annotations are intended to train models to understand and predict subsequent actions in a sequence, a crucial skill for automating RPA tasks across different web environments. Essentially, GUIDE pays homage to and builds upon the work seen in datasets like UiPath's Public Dataset, blending real-world RPA challenges with the advanced capabilities found in models such as PixelBERT and MuRIL. As we enter an era where LLMs are increasingly relied upon to comprehend and navigate the digital realm akin to humans, GUIDE emerges as a guiding light for research, offering the potential to unlock new horizons in automation, natural language processing, and beyond. This paves the way for a future where AI seamlessly integrates into the fabric of everyday digital tasks.

## 3. Stages of Data Collection

At a high level, the process contains of four stages (i) Pre-Data Collection (ii) NEXTAG (iii) Quality Check and (iv) Post processing. We describe each of these in detail below.

### 3.1. Pre-Data Collection

#### 3.1.1. Tasks Collection

In the initial phase of our data collection process, we actively gather a wide range of tasks from various customers. These tasks reflect real-world scenarios and actions that individuals and organizations might require automation for. The focus is on ensuring diversity in task complexity and industry domains. The collection is accomplished through multiple channels to capture a broad spectrum of needs, such as: Direct submissions from business entities detailing their routine operations, Survey responses from individual professionals highlighting frequent tasks they wish to automate. During this stage, we also emphasise on the quality and clarity of the requests. Each task is documented with enough detail to allow for accurate interpretation and subsequent automation.

#### 3.1.2. Task Filtering

Post-collection, we implement a rigorous filtering process to ensure that the tasks are suitable for RPA[13] and will provide valuable data for the GUIDE dataset. The criteria for filtering include: Feasibility: Tasks should be executable through RPA technology. Legality: Tasks should comply with legal regu-lations and ethical standards. Clarity: Tasks should be well-defined without ambiguity. Tasks are reviewed by our team to assess their fit into our criteria. Those that do not meet the quality thresholds are removed from the collection. This filtering stage plays a critical role in maintaining a high standard for our dataset and ensures relevance to RPA development needs.

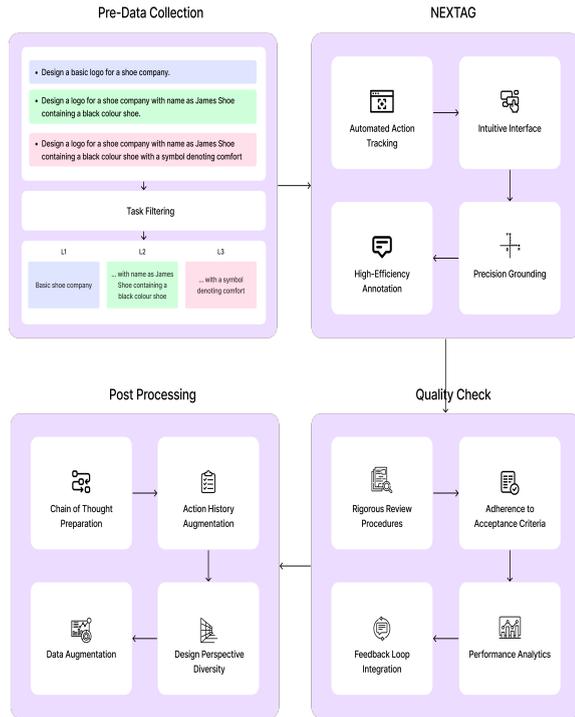

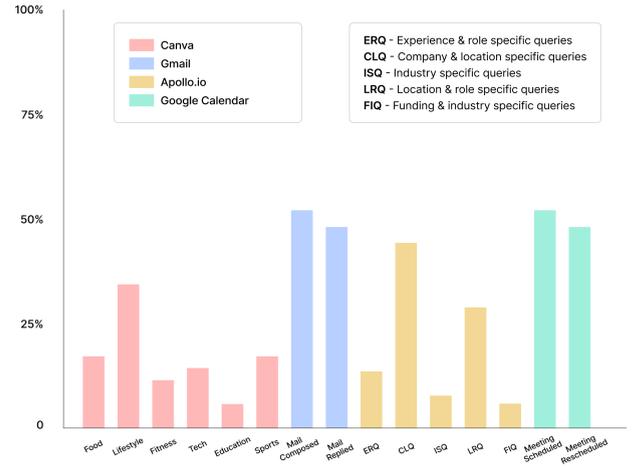

Figure 1: *Four Stages of Data Collection*

Figure 2: *Domains of data collected*

### 3.1.3. Hierarchical Categorization

The tasks passing the filtering stage are then categorized into three distinct levels according to their complexity, specificity, and the extent to which they require advanced cognitive processing:

- Level 1: Basic tasks that involve straightforward, single-step actions.
- Level 2: Intermediate tasks that incorporate multiple steps and require moderate decision-making.
- Level 3: Complex tasks that demand sophisticated decision-making, multi-step sequences, and context-awareness.

Categorizing tasks allows us to structure the GUIDE dataset in a way that facilitates targeted model training. It helps in highlighting the progressive learning needs for RPA models as they ascend from executing simple tasks to handling intricate processes. This structured approach aids in evaluating the performance and adaptability of LLM models with varying levels of task complexities. Additionally, tasks within each category are annotated with insights reflecting the rationale behind their categorization. This provides additional metadata for researchers and developers to understand and exploit the dataset more effectively, tailoring their AI models to the nuances across different levels of task complexity.

### 3.2. NEXTAG

The NEXTAG (Next Action Grounding and Annotation Tool) is an innovative in-house tool designed to streamline the data annotation process for the GUIDE dataset. It represents a breakthrough in the way data for Robotic Process Automation (RPA) tasks is annotated, offering a high degree of efficiency and accuracy in capturing user interaction data.

#### 3.2.1. Automated Action Tracking

NEXTAG significantly reduces manual labor by automatically recording all actions taken by a user within a web browser. As users navigate through tasks, NEXTAG meticulously logs each click, scroll, and text input, ensuring comprehensive data capture without disrupting the natural flow of task execution.

#### 3.2.2. Intuitive Interface

The tool is equipped with a user-friendly interface that simplifies the navigation and interaction process. This promotes ease of use and allows annotators to focus on task execution rather than the complexities of manual data entry.

#### 3.2.3. High-Efficiency Annotation

Annotation speed is exponentially increased with NEXTAG's automated recording capabilities. It removes the need for annotators to manually document successive actions and corresponding coordinates, drastically reducing the time taken for data preparation.

#### 3.2.4. Precision Grounding

NEXTAG precisely identifies and records the spatial coordinates of GUI elements where user actions occur. This granularity is crucial for training RPA models that must execute tasks by interacting with user interface components.

It provides consistent and error-free annotations, thereby ensuring high-quality data that is essential for the accurate training of RPA models. NEXTAG's efficiency enables scaling the dataset collection process to incorporate a wider variety of tasks, applications, and user interactions. The richness of data captured by NEXTAG facilitates in-depth analysis and the development of models with enhanced understanding and predictive capabilities for complex task sequences. NEXTAG epitomizes the advancement in data annotation tools by combining

**Algorithm 1** Guided Exploration within the NEXTTAG Framework

**Require:** The 'World Interface' $W$, an abstract representation of the OS environment
**Ensure:** Transformed action steps $A$, aligned with corresponding events $E$ in the environment
1: Initialize Event Set $E \leftarrow \{\}$
2: Initialize Action Set $A \leftarrow \{\}$
3: **for** each interface element $i$ in $W$ **do**
4:     Capture event $e_i$ with its metadata:
5:         timestamp $\leftarrow$ GetCurrentTimestamp()
6:         action_type $\leftarrow$ GetActionType($e_i$)
7:         action_description $\leftarrow$ GetActionDescription($e_i$)
8:         grounding_description $\leftarrow$ GetGroundingDesc($e_i$)
9:     Add event metadata to $E$:
10:        $E \leftarrow E \cup \{e_i(\text{metaData})\}$
11:     Apply transformation function $\tau_{\text{explore}}$ to record actions:
12:        $a_i \leftarrow \tau_{\text{explore}}(e_i)$
13:     Add transformed action to $A$:
14:        $A \leftarrow A \cup \{a_i\}$
15:     Check if $a_i$ leads to a new state in $W$
16:     **if** a new state is reached **then**
17:        Update $W$ with the new state information
18:     **end if**
19: **end for**
20: **PostProcessing:**
21: **for** each recorded action $a_i$ in $A$ **do**
22:     Process and Augment $a_i$ ( Grounding, ImageAugmentation, Scaling)
23:     **if** user modifies $a_i$ **then**
24:        Record user's modified action $a'_i$
25:        Update $A$ with $a'_i$:
26:           $A \leftarrow (A \setminus \{a_i\}) \cup \{a'_i\}$
27:     **end if**
28: **end for**
29: **return** $A, E$

efficiency with precision. It paves the way for robust training datasets like GUIDE, which are instrumental in the evolution of RPA models. With NEXTTAG's capabilities, we set a new standard in data annotation, effectively bridging the gap between human-like understanding and automated task execution in the domain of RPA.

### 3.3. Quality Check

To maintain the integrity and utility of the GUIDE dataset, our data collection process incorporates a critical Quality Check (QC) stage managed by an in-house Quality Control team. This team is tasked with the careful examination of every entry within the dataset, including images and annotations, to verify their accuracy and consistency with our high-quality standards. The QC team functions as a gatekeeper, guaranteeing that the data fed into training models is free from errors and aligned with the objectives of precision in RPA tasks. Here's an overview of the Quality Assurance process

#### 3.3.1. Rigorous Review Procedures

Our QC team employs a comprehensive review procedure that includes a point-by-point inspection of each dataset entry. This process is designed to detect and correct any inaccuracies in the annotations and to confirm that the captured actions are detailed precisely in alignment with the corresponding GUI elements.

#### 3.3.2. Adherence to Acceptance Criteria

We have established stringent acceptance criteria that every data point must meet before it becomes part of the GUIDE dataset. These criteria encompass aspects such as the clarity of images, the correctness of annotated actions and grounding coordinates, as well as the relevance and feasibility of the actions for RPA tasks.

#### 3.3.3. Feedback Loop Integration

The Quality Check process isn't just about identifying issues; it's also about learning from them. Our QC team provides feedback to the annotators and tool developers, guiding them on areas for improvement and contributing to the continuous refinement of NEXTAG and the overall data collection process.

#### 3.3.4. Performance Analytics

We utilize performance analytics to track the effectiveness of our QC measures, setting quantifiable goals for accuracy and using metrics to guide enhancements in our QC methodologies.

With a robust Quality Check system, we fortify the accuracy and relevance of the GUIDE dataset, enabling RPA models to be trained on reliable data and perform at their best when automating real-world tasks. This foundational quality assurance is pivotal in ensuring that the outcomes of machine learning algorithms are dependable and that they can seamlessly translate into practical, automated solutions in complex GUI environments.

### 3.4. Post Processing

Upon the completion of data collection and the implementation of rigorous quality checks, the GUIDE dataset undergoes a critical post-processing phase. This stage is essential for preparing the data to train RPA systems that are capable of operating across varied environments and understand the context of each task. The post-processing encompasses several key components:

#### 3.4.1. Chain of Thought (CoT) Preparation

We enrich the dataset with a Chain of Thought (CoT) for each task, which provides a narrative representing the internal reasoning a model might undertake to arrive at the next action. This CoT data is carefully constructed to reflect logical progression and rationale, simulating human-like problem-solving sequences that can be instrumental for training more insightful and interpretable RPA models.

#### 3.4.2. Action History Augmentation

To further contextualize the task at hand, the dataset includes a history of previous actions leading up to the current state of the task flow. This historical context is appended to each entry, allowing RPA models to understand the continuity of tasks and make informed decisions about subsequent actions.

#### 3.4.3. Data Augmentation

The dataset undergoes several augmentation techniques to ensure that models are trained on data that reflects the diverse conditions under which RPA tasks might occur. Various augmentation techniques are applied such as Images with multiple browsers, multiple Operating Systems, Entries with both dark and light mode aesthetics, Special augmentation techniques, such as applying borders, cropping, and shifting of interface elements, help in creating a dynamic grounding dataset. These modifications prevent the model from simply memorizing coordinates by encouraging the understanding of element positioning relative to other interface features.

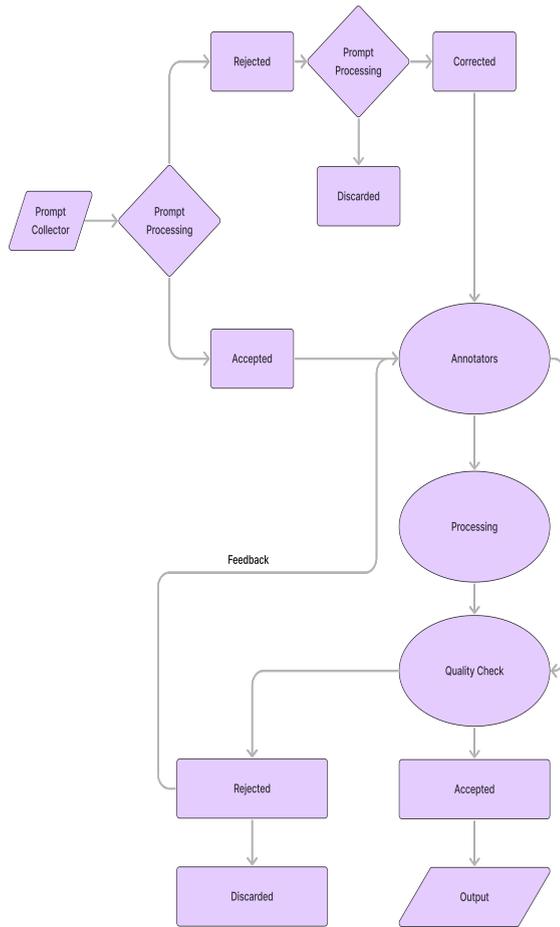

Figure 3: *Data collection Pipeline*

*3.4.4. Design Perspective Diversity*

Annotators from various design backgrounds contribute to the augmentation process, ensuring that the dataset captures a wider range of design aesthetics and user interaction patterns. This diversity is valuable in training models that are resilient to changes in website design or layout.

The comprehensive post-processing of the GUIDE dataset is instrumental in building models that can generalize from the training data to real-world situations. By incorporating diverse representations of browser and OS environments, and by simulating various visual themes and design paradigms, we provide a rich training ground that challenges models to learn the essence of GUI element recognition and action grounding. This multiplicity ensures that RPA models can go beyond rote memorization of screen coordinates and develop a more sophisticated understanding of task execution in any given interface, thus significantly enhancing their real-world applicability.

## 4. Task Difficulty

In striving to create RPA models that can adeptly handle tasks of varying intricacy, the GUIDE dataset incorporates a meticulous Task Complexity Analysis. This structure organizes tasks into a hierarchical categorization, delineating them into discrete levels based on their difficulty. Here, we provide a comprehensive look at the methodology behind this analysis and the criteria for each difficulty level.

### 4.1. Task Complexity Analysis

The primary goal of classifying tasks by complexity level within the GUIDE dataset is to enable the creation of RPA systems that can intelligently scale their problem-solving strategies to match the task at hand. Provide a scaffold for incremental learning, where models can first master simpler tasks before progressing to more complex scenarios. Facilitate a benchmarking framework that can assess RPA performance across a gradient of challenges, from the most basic to the highly intricate.

### 4.2. Categorization Criteria

Tasks within GUIDE are subjected to a thorough evaluation based on

- Task Singularity: The number of discrete actions required to complete the task. Simpler tasks are often few-step(1-5), while more complex ones involve a sequence of actions.

- Cognitive Load: The level of decision-making, problem-solving, and contextual reasoning necessary to successfully perform the task.

- Data Integration: The extent to which a task requires synthesizing information from multiple sources or across different timeframes within a workflow.

- User Interface Dynamics: The variability in the GUI layout, such as the presence of dynamic elements like drop-downs, modals, or custom form fields which add complexity.

- Exception Handling: The potential for unpredictable events, requiring the model to navigate errors or unexpected changes in the interface.

### 4.3. Description of Difficulty Levels

- Level 1(Low Complexity): Comprised of tasks that are straightforward, often involving direct interactions like clicking a clearly labeled button or inputting text into a designated field. Minimal decision-making is required, with no conditional logic or significant contextual information to consider.

- Level 2(Moderate Complexity): Includes tasks that necessitate multiple steps or interactions, such as filling out a form or navigating through a sequence of pages to retrieve information. Moderate decision-making and informational processing are involved, with some reliance on the task's context.

- Level 3(High Complexity): Encompasses tasks that are multifaceted, usually comprising complex workflows like data analysis across multiple systems or managing intricate user interactions. High cognitive load with substantial decision-making demands, involving dynamic problem-solving under varied conditions and exceptions.

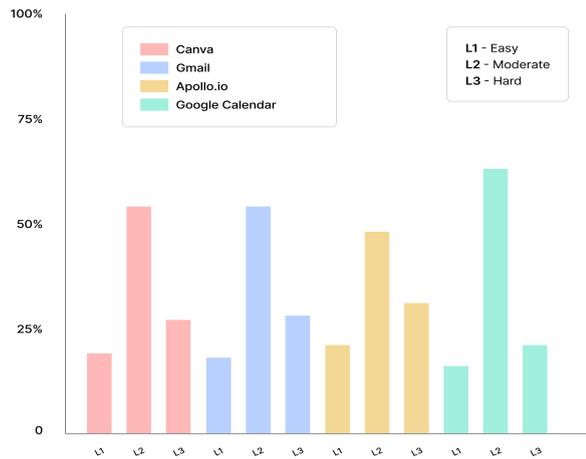

Figure 4: *Categorization of data into different levels*

## 4.4. Impact on Model Training and Evaluation

The layout of these complexity levels supports a tailored approach to model training. By starting with basic tasks, models can establish foundational skills in GUI interaction before embracing more challenging scenarios. Progression through complexity levels allows for a staged training regimen, much like structured learning curricula in human education. Furthermore, the categorization facilitates nuanced evaluation, enabling us to gauge a model's proficiency across a spectrum of task difficulties and to identify specific areas where additional refinement is needed. In summary, the Task Complexity Analysis of the GUIDE dataset serves as a backbone for building RPA models that are responsive to the richness and variability inherent to real-world automation tasks. It allows for a strategic training process that not only aims for breadth and depth of understanding but supports the continuous evolution of AI capabilities in the vibrant landscape of RPA challenges.

## 5. Data Augmentation

Data augmentation is a crucial technique applied in machine learning to expand the original dataset by generating modified versions of existing data or creating new synthetic data from existing data. This process is particularly valuable when a dataset has limited samples or lacks diversity. Augmentation helps to avoid model overfitting—where a model learns details and noise in the training data to the extent that it negatively impacts the performance of the model on new data and improves the model's generalization capabilities across various conditions and environments. In the context of the GUIDE dataset and RPA, data augmentation allows us to simulate a more comprehensive range of scenarios that an AI model may encounter when interacting with Graphical User Interfaces (GUIs) in real-world situations.

### 5.1. Forms of Data Augmentation

#### 5.1.1. Browser Diversity

Models are trained to recognize and interact with GUI elements within web pages. By simulating how these pages might render on different browsers (e.g., Chrome, Firefox, Safari), we prepare the model to accurately perform tasks irrespective of the end-user's choice of browser. Images within the dataset have additional tabs or interface elements superimposed to represent different web browsers, encouraging the model to recognize and adapt to various browser-specific GUI layouts.

#### 5.1.2. Operating System Variability

Interfaces can significantly vary based on the operating system. Augmenting data with images that include different OS-specific GUI elements (like window decorations, system menus, and icons) enhances the model's adaptability to platforms like Windows, macOS, and various Linux distributions. Dataset images also include menus or visual cues characteristic of different operating systems, ensuring the model's familiarity with navigating diverse OS interfaces.

#### 5.1.3. Theme Adaptation

With the prevalent use of dark and light modes in applications, models need to recognize and interact with elements across both themes. Augmentation provides images where GUI components are in these different visual modes, equipping models to maintain performance regardless of the interface theme.

#### 5.1.4. Spatial Variations

GUI elements may not always be in the same location (due to responsive design or user customization). Models must understand the relative positioning of elements rather than memorize fixed coordinates. Techniques like adding borders, cropping, or shifting elements help create a dataset that accounts for this spatial variability.

### 5.2. Impact on RPA Model Performance

#### 5.2.1. Robustness

By training on augmented data, RPA models develop a robust understanding of the GUI. This means that they become less sensitive to changes in appearance or layout, thereby reducing the risk of failure when deployed in changing digital environments.

#### 5.2.2. Generalization

Models exposed to a diverse set of images and scenarios can generalize better. They're more likely to accurately interpret and act upon previously unseen interfaces or unexpected changes, hence improving their overall reliability.

#### 5.2.3. Flexibility

Data augmentation aids in developing flexible models that are not tied to a singular application design or environment, making them more practical for varied end-users with different preferences and settings.

#### 5.2.4. Error Reduction

Training on augmented data helps models recognize the salient features of GUI elements, thereby reducing the likelihood of misclicks or incorrect navigations, which are critical in ensuring successful automation task completion.

Implementing a rigorous and thoughtful data augmentation strategy enhances the GUIDE dataset's capability to train RPA models that are well-equipped for real-world tasks. Incorporating varied GUI representations bolsters models' resilience against overfitting and prepares them for the inevitable variability they will encounter in practical applications. The resulting sophistication and reliability of these models have profound implications for the scale at which RPA can be implemented across industries and applications.

# 6. Limitations

Despite GUIDE's comprehensive approach to data collection and annotation for RPA tasks, there are inherent limitations that should be acknowledged. Understanding these limitations is essential for setting realistic expectations and guiding future developments of the dataset and associated models. Here are some of the potential limitations of the GUIDE data

### 6.1. Limited Domain Scope

Although GUIDE covers a range of websites and tasks, it may not encapsulate the full spectrum of domains and industries that could benefit from RPA. Automation needs in sectors not represented in the dataset might not be addressed adequately.

### 6.2. Annotation Bias

The data annotations are subject to the interpretations and judgments of the annotators, which can introduce bias. Despite efforts to standardize annotations, variations in understanding complex tasks or interface layouts might affect the dataset's consistency.

### 6.3. Interface Dynamics and Updates

Web interfaces are fluid and subject to frequent changes. GUIDE data collected at a particular point may not reflect these updates, which could affect an RPA model's performance when dealing with the latest interface versions.

### 6.4. Simulated Environment Limitations

While data augmentation techniques such as adding elements to represent different operating systems and browsers are beneficial, they cannot entirely replicate the nuances of genuinely operating in those environments. Subtle differences might not be accounted for in the dataset.

### 6.5. Real-World Interaction Complexity

GUIDE dataset simulations may not capture the complexity of human interactions with GUIs in real-world scenarios fully. Users may employ keyboard shortcuts, right-click contextual menus, or other advanced interactions that are beyond the scope of the dataset.

### 6.6. Exception and Error Handling

Automated tasks often encounter exceptions or errors that require human judgment. GUIDE data might not offer sufficient examples of these edge cases, limiting models' ability to handle unexpected scenarios robustly.

### 6.7. Scalability of Data Collection

NEXTAG streamlines the data annotation process significantly, but the scalability of data collection could still be constrained by the need for manual navigation and the availability of diverse annotators to execute tasks.

The GUIDE dataset is an invaluable resource for advancing the capabilities of RPA models, yet it is crucial to continue refining the dataset and developing methodologies to address these limitations. Future iterations may benefit from expanding the domain coverage, enhancing the realism of simulated environments, and incorporating more comprehensive data on exception handling to ensure models trained on GUIDE can be effectively deployed in a wide variety of real-world scenarios.

# 7. Experimental Overview

Our evaluation demonstrates that V-Zen effectively grounds query instructions to the corresponding GUI elements across various software platforms. Through a series of grounding experiments, we observed a remarkable improvement in the model's ability to discern and interact with GUI components, such as buttons, fields, and tabs, in comparison to baseline methods. This is attributed to the model's employment of high cross-resolution visual processing and deep fusion of visual-language features.

### 7.1. Quantitaive Analysis

Comparing the performance of GPT-4 Vision, Gemini Pro, and our model V-Zen across Next Task Prediction and Grounding tasks reveals that it surpassed some strong baselines with seemingly less data. GPT-4 Vision showcases a slightly higher accuracy of 94% in predicting the next task, attributed to its advanced architecture and comprehensive pre-training on vision tasks. However, V-Zen closely follows with an accuracy of 93.2%, indicating its competitive capability in task prediction. In grounding tasks, GPT-4 Vision achieves a accuracy of 28%, while Gemini Pro notably lags behind with only 21%. Surpassing both, V-Zen demonstrates an impressive grounding accuracy of 89.7%, highlighting its adeptness in understanding task contexts and environments. These results underscore V-Zen's potential for real-world applications requiring precise task execution and contextual understanding, emphasizing the importance of robust grounding capabilities alongside accurate task prediction in AI model development. This indicates that our LVLM (Language-Vision Language Model) possesses strong reasoning and planning capabilities, surpassing the state-of-the-art (SOTA) visual methods. V-Zen particularly stood out in tasks involving dynamic interface layouts and varying element positions, confirming our hypothesis that conventional LVLMs face challenges in accurately interacting with graphical user interfaces.

Table 1: *Performance of the proposed model with different open sourced datasets.*

|  | Next Task Prediction | Grounding |
|---|---|---|
| GPT-4V | 94 | 28 |
| Gemini-Pro | 92 | 21 |
| V-Zen | 93.2 | 89.7 |

### 7.2. Grounding Ablation Study

The table presents the outcomes of an ablation study concerning Grounding accuracy, examining the impact of various components on model performance. Starting with the baseline of Raw Data, the study incrementally incorporates different elements, including Previous Action, Previous Action History, Chain of Thought, Augmentation (comprising crop, border, noise, jitter, and shifting), and OS/Browser information. Notably, the accuracy improves with each additional component, indicating the significance of each aspect in enhancing the model's ground-ing capabilities. The most substantial improvement is observed when incorporating various augmentation techniques on images and adding OS/Browser information to handle across all platforms, leading to a notable accuracy of 78.3%, showcasing the crucial role of image augmentation and diversity in refining grounding accuracy. We can conclude that the accuracy improvement due to augmentation adds to model's robustness to unseen data and reduces its sensitivity to variations. The generalization capabilities of the model also improves across a range of features and patterns present in the data and it helps address data imbalance. In our study we have systematically introduced components and divides annotations into 'input' and 'output' groups. Input information includes screen descriptions and previous action history results, while action thinking capabilities and action description are incorporated into the target output to facilitate understanding the chain-of-thought thinking process. Notably, integrating previous action results enhances the coherence of cognitive decision making, resulting in accuracy improvements. This underscores the importance of such semantic annotations in refining model performance. We have also calculated grounding accuracy provided next task prediction response given is correct. By incremental increase, finally our model is able to achieve an overall accuracy of 89.7%.

Table 2: *Ablation Study wrt Grounding.*

| Method | Accuracy |
|---|---|
| Raw Data | 60.8 |
| *+Previous Action | 64.2 |
| *+Previous Action History | 65.6 |
| *+Chain of Thought | 63.7 |
| *+Augmentation[crop, border, crop+broder, noise, jitter, shifting] | 78.3 |
| *+OS/Browser | **89.7** |

### 7.3. Next Task Prediction Ablation Study

The outcomes of an ablation study focusing on Next Task Prediction accuracy, outlining the impact of various components on model performance were thoroughly studied. Beginning with the baseline of Raw Data, we studied the steady increase in the accuracy. Here, the introduction of Chain of Thought leads to a substantial improvement, boosting accuracy to 92.4%. This indicates the crucial role of incorporating logical progression and reasoning sequences in enhancing the model's ability to predict subsequent tasks effectively. Moreover, the inclusion of Augmentation techniques results in further performance improvement, elevating accuracy to 93.7%. Thus CoT facilitates a more deeper understanding of user intent and interaction dynamics by conducting a thorough analysis of screen context, action reasoning, targets, and results and adding that in the output segment of our data. The inclusion of the previous action result addresses the issue of estimating task progress by clearly defining the outcome of past actions in textual form, thus establishing a link between the past and present. This shows the significance of maintaining the continuity of action decisions by establishing connections between consecutive execution steps.

Table 3: *Ablation Study wrt Next Task Prediction.*

| Method | Accuracy |
|---|---|
| Raw Data | 59.5 |
| *+Previous Action | 67.5 |
| *+Previous Action History | 78.2 |
| *+Chain of Thought | 92.4 |
| *+Augmentation[crop, border, noise, jitter, shifting] | 93.7 |
| *+OS/Browser | **93.2** |

The another enhancement is observed when incorporating OS/Browser information, achieving an accuracy of 93.2%.

## 8. Conclusion

In conclusion, the GUIDE dataset marks a substantial advancement in the realm of Robotic Process Automation (RPA), enriching the field with a robust resource for training AI models on next-action prediction and action grounding tasks. With its unique combination of detailed task descriptions, annotated actions, and varied augmentation strategies, GUIDE addresses the critical need for datasets that encapsulate the complexity of real-world graphical user interfaces. While GUIDE has set a precedent for quality and comprehensiveness in RPA datasets, we acknowledge its limitations, including the scope of task domains and the challenge of capturing the full dynamism of web interfaces. Nonetheless, the dataset offers a promising foundation for developing intelligent RPA systems capable of navigating the nuances of GUI-based tasks with greater accuracy and adaptability. As the community continues to utilize and build upon the GUIDE dataset, we anticipate a future where AI-driven automation transcends the limitations of static scripts and embraces the fluid, context-aware interactions that characterize human computer usage. The GUIDE dataset stands as both a milestone and a catalyst for ongoing innovation in the automation of complex task sequences within diverse digital environments.

# 10. Appendix

## 1. GitHub

The data repository, along with detailed documentation, can be found at `https://github.com/superagi/GUIDE.`

## 2. Huggingface

The GUIDE data, along with detailed documentation is available at `https://huggingface.co/datasets/SuperAGI/ GUIDE`

## 3. Qualitative Examples

**Prompt**

Task: Write an email to Kevin at kevin@gmail.com asking him about the update on the multimodal model.Previous Action: TYPE: Type the email content in the content box button Give me the next action?

**Response**

Reasoning: After writing the content of the mail, the next action is to click on the send button. CLICK: Click on the send [0.804134213, 0.664244255, 0.9642443221, 0.978234231] button

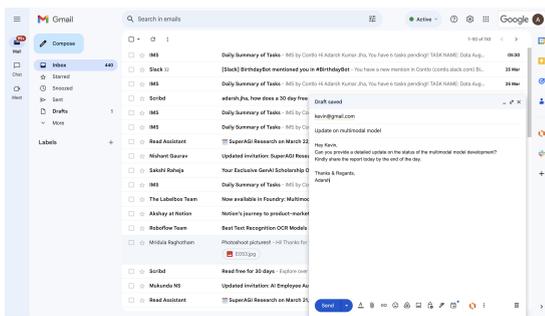

Figure 5: *Gmail*

**Prompt**

Task: Give me the list of all product manager with minimum 5 years of experience working in a series A company with minimum 1M in revenue. Previous Action: TYPE: Type Job Titles in the Search filters tab. Give me the next action?

**Response**

Reasoning: After typing Job Titles in the more filters tab, the next action is to click on the Job Titles button. CLICK: Click on the Job Titles [0.184375, 0.5018518518518519, 0.38958333333333334, 0.5657407407407408] button

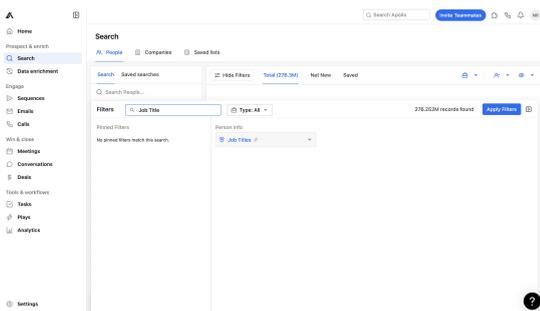

Figure 6: *Apollo.io*

**Prompt**

Task: Financial Planning App Logo Design: Create a logo for WealthWise Finance App a mobile application providing users with tools and resources for budgeting, investing, and financial planning. Utilize elements like financial charts, savings symbols, or a smartphone icon. Download in SVG format. Previous Action: DOUBLE CLICK: Double click on the bounded text box tab. Give me the next action?

**Response**

Reasoning: After Double clicking on the bounded text box tab, the next action is to type WealthWise in the bounded text box tab..TYPE: Type WealthWise in the bounded text box [0.5171875, 0.6296296296296297, 0.7666666666666667, 0.6907407407407408] tab.

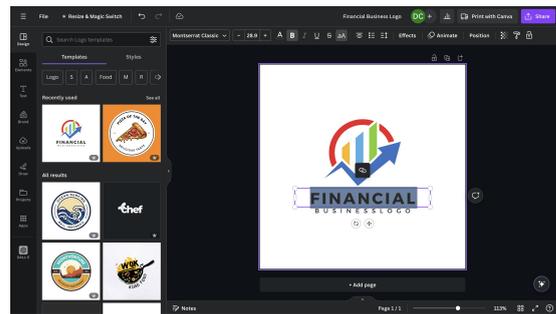

Figure 7: *Canva*

**Prompt**

Task: Schedule a meeting with Devin at 6 p.m on 24th March regarding the multiagent updates. Previous Action: Type: Type devin@gmail.com in the bounded text box tab. Give me the next action?

**Response**

Reasoning: After adding the guests of the meeting, the next action is to click on the send button.CLICK: Click on the send [0.6098958333333333, 0.5824074074074074, 0.6427083333333333, 0.6129629629629629] button.

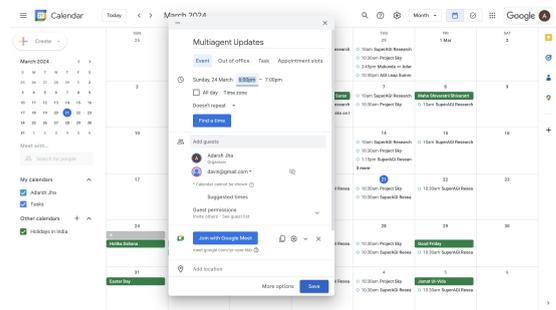

Figure 8: *Google Calendar*